\documentclass{article}

\usepackage{PRIMEarxiv}

\usepackage{graphicx}
\usepackage{epstopdf}
\usepackage{natbib}
\usepackage{amsmath,amsfonts,amssymb}
\usepackage[utf8]{inputenc} 
\usepackage[T1]{fontenc}    
\usepackage{hyperref}       
\usepackage{url}            
\usepackage{booktabs}       
\usepackage{amsfonts}       
\usepackage{nicefrac}       
\usepackage{microtype}      
\usepackage{lipsum}
\usepackage{fancyhdr}       

\newcommand{\e}{\mathrm{e}}
\newcommand{\upd}{\mathrm{d}}
\newcommand{\upi}{\mathrm{i}}

\title{Simultaneous layout and device parameter optimisation of a wave energy park in an irregular sea
}

\author{\textbf{Ben Wilks$^{1*}$, Michael H. Meylan$^{1}$, Fabien Montiel$^{2}$, Dasun Shalila Balasooriya$^{1}$, Tahir Jauhar$^{3}$,\\ Craig Wheeler$^{3}$ \& Stephan Chalup$^{1}$}\\
\begin{enumerate}
    \item School of Information and Physical Sciences, University of Newcastle, Callaghan NSW 2308, Australia
    \item Department of Mathematics and Statistics, University of Otago, Dunedin 9016, New Zealand
    \item School of Engineering, University of Newcastle, Callaghan NSW 2308, Australia
\end{enumerate}
* Email for correspondence: ben.wilks@newcastle.edu.au}

\begin{document}
\maketitle

\begin{abstract}
The design of optimal wave energy parks, namely, arrays of devices known as wave energy converters (WECs) that extract energy from water waves, is an important consideration for the renewable transition. In this paper, the problem of simultaneously optimising the layout and device parameters of a wave energy park is considered within the framework of linear water wave theory. Each WEC is modelled as a heaving truncated cylinder coupled to a spring-damper power take-off. The single-WEC scattering problem is solved using an integral equation/Galerkin method, and interactions between the WECs are solved via a self-consistent multiple scattering theory. The layout of the array and power take-off parameters of its constituent devices are simultaneously optimised using a genetic algorithm, with the goal of maximising energy absorption under a unidirectional, irregular sea described by a Pierson--Moskowitz spectrum. When constrained to a rectangular bounding box that is elongated in the direction of wave propagation, the optimal arrays consist of graded pseudo-line arrays when the number of WECs is sufficiently large. Moreover, low-frequency waves propagate further into the array than high-frequency waves, which is indicative of rainbow absorption, namely, the effect wherein waves spatially separate in a graded array based on their frequency, and are preferentially absorbed at these locations. Arrays optimised for a square bounding box did not show strong evidence of grading or rainbow reflection, which indicates that more complicated interaction effects are present.
\end{abstract}

\section{Introduction}
The development of technologies to harness the energy of ocean waves is an important avenue of research for the renewable transition, as the large amount of energy contained in ocean waves makes them an attractive resource \citep{guo2021}. To date, the existing technology for ocean wave energy conversion is not as economically efficient as wind or solar energy conversion, preventing widespread adoption. To address this, the main approach has been to improve the design of energy-harnessing devices, commonly referred to as wave energy converters \citep[WECs,][]{FALNES2007185,FALCAO2010899,SHENG2019482}. This paper considers a secondary approach, which is to optimise the configurations of arrays of WEC, known as wave energy parks \citep{goteman2020advances,TEIXEIRADUARTE2022112513,GOLBAZ202215446}.

Optimisation of wave energy parks is a challenging computational task. Simulating wave interactions with the park is expensive, and optimisation becomes more expensive as the number of WECs increases, as the number of WECs determines the number of variables to be optimised. Previous approaches have simplified the task by fixing the device parameters during optimisation, allowing only the park layout to vary. \citet{CHILD20101402} optimised the layout of an array of cylindrical WECs for maximum power take-off in a regular sea, using both a genetic algorithm and a parabolic intersection method. The parameters of the device were optimised prior to the layout optimisation. Using the same algorithms, \citet{mcguinness2018hydrodynamic} considered several layout optimisation problems for arrays of point absorbers, including line arrays, circular arrays, and arbitrary layouts. The more realistic problem of optimising the array for an irregular sea was considered by \citet{child2011configuration} and \citet{mcguinness2018hydrodynamic}, in which the sea state was described by the JONSWAP spectrum in both studies. \cite{GIASSI2018252} used a genetic algorithm to optimise an array for absorption from an irregular sea described by time-series data from a site in Sweden. \citet{NESHAT2020100744} considered the problem of optimising an array of submerged three-tether buoys in irregular seas described using observational data taken from sites near Australian cities. In this study, the directional spectrum of the incident waves was also considered. The problem of optimising for irregular sea conditions is more computationally challenging because the underlying wave-park interaction model has to be evaluated across a range of frequencies, making objective function evaluation more expensive. Our task in this paper is to simultaneously optimise the layout and device parameters of the park---we are not aware of any similar work. In particular, the parameters of the park will be tuned in order to maximise power take-off from an irregular sea. In addition to maximising power take-off, previous work has considered secondary objectives, such as minimising installation and maintenance costs of the associated electrical infrastructure of the park \citep{sergiienko2018,bergstrom2024comprehensive}, using multiple objective optimisation. This is outside the scope of the current paper.

A promising concept to advance wave energy park optimisation is rainbow absorption. This phenomenon, which arises from the field of metamaterials, occurs when waves propagate through an array with graded, locally resonant properties. The graded structure causes the wave energy to gradually slow down and amplify in a location that depends on its frequency, before being absorbed through a loss mechanism. Following earlier investigations in the context of acoustics \citep{jimenez2017} and elasticity \citep{Chaplain_2020}, rainbow absorption has since been studied in water waves. \citet{Wilks2022} considered a device consisting of multiple surface-piercing vertical barriers with a graded submergence depth in a two-dimensional fluid, which slows down and spatially separates incident energy into different regions. Rainbow absorption was induced by a loss mechanism consisting of rectangular pistons between each adjacent pair of barriers, which were coupled to a linear power take-off system. When optimised using a local search algorithm, the resulting device achieved near-perfect energy absorption over a prescribed frequency interval (more than $98\%$ of the energy absorbed). \citet{Westcott_Bennetts_Sergiienko_Cazzolato_2024} achieved near-perfect energy absorption by an array of rectangular WECs in the absence of a surrounding structure (i.e. barriers) by using springs to tune the resonances of the WECs and therefore the grading of the array. The WEC parameters that achieved this were found using a two-stage local search optimisation procedure, by first tuning the spring stiffness coefficients to achieve near-perfect reflection by the array and then tuning the power take-off parameters for near-perfect absorption. In these aforementioned studies of rainbow absorption, the grading of the device was imposed as a constraint in the optimisation procedure. While this assumption is sufficient to achieve near-perfect energy absorption in a two-dimensional context, it is unclear whether (and under what conditions) grading would be optimal in a more realistic three-dimensional model. This paper addresses this question by optimising a three-dimensional wave energy farm in the absence of any \textit{a priori} assumption of grading.

The outline of this paper is as follows. In \textsection\ref{Model_and_solution_sec} we introduce a model of a WEC that consists of a truncated, partially submerged cylinder, which is constrained to move in heave and coupled to a spring-damper system. Solutions to the single- and multiple-scattering problems are then described, and details of the array power take-off calculations used later in the paper are given. In \textsection\ref{optimisation_sec}, the constrained optimisation problem for the array is stated, in which both the coordinates and the spring-damper parameters of the WECs are variables. Optimally configured arrays and their absorption spectra, which we find using a genetic algorithm, are given in \textsection\ref{results_sec}. A discussion is given in \textsection\ref{discussion_sec}.

\section{Problem formulation and solution}\label{Model_and_solution_sec}
\subsection{Scattering by a single WEC}
\begin{figure}
    \centering
    \includegraphics[width=0.7\linewidth]{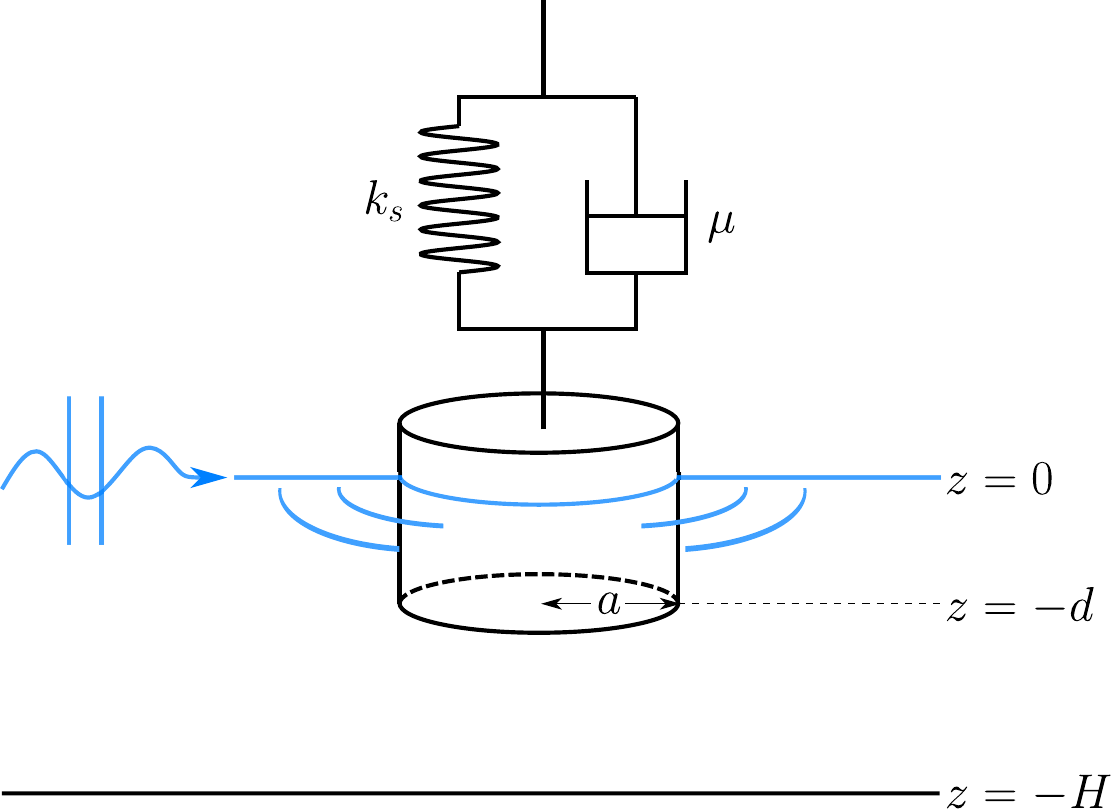}
    \caption{Schematic of the cylindrical WEC model considered in this paper. The radius and equilibrium submergence of the cylinder are denoted $a$ and $d$, respectively, and the fluid is of constant depth $H$. An incident wave excites the WEC into vertical motion, which drives an external spring-damper system (with spring and damping coefficients $k_s$ and $\mu$, respectively) and also generates scattered waves in the fluid.}
    \label{fig:schematic}
\end{figure}

We initially consider a single WEC consisting of a truncated cylinder partially submerged in a fluid of constant depth $H$. The WEC is constrained to move in heave and coupled to a power take-off mechanism that we model as a linear spring-damper system. The problem is initially posed in a Cartesian coordinate system, in which the $xy$-plane coincides with the undisturbed free surface of the fluid and the $z$ axis, which points vertically upwards, coincides with the central axis of the WEC. With the intention of applying the method of separation of variables, the system is transformed into cylindrical coordinates $(r,\theta,z)$ in which $x=r\cos\theta$ and $y=r\sin\theta$. A scattering problem in the fluid is posed using time-harmonic linear water wave theory, which assumes that the fluid is incompressible and inviscid and undergoing irrotational, time-harmonic motion with time dependence $\e^{-\upi\omega t}$ \citep{Linton2001,Mei2005}. Under these assumptions, the problem reduces to finding the complex potential $\phi$, which satisfies the following boundary value problem:
\begin{subequations}\label{BVP}
\begin{align}
    \bigtriangleup \phi&=0& (r,\theta,z)\in\Omega\label{laplace}\\
    \partial_z\phi&=-\upi\omega s&r<a,z=-d\label{piston_dyn_BC}\\
    \partial_z\phi&=\frac{\omega^2}{g}\phi &r>a,z=0\label{Free_surface}\\
    \partial_z\phi&=0&z=-H\label{seabed}\\
    \partial_r\phi&=0&r=a,z>-d,\label{barrel}
\end{align}
\end{subequations}
where $\Omega$ is the fluid domain, $g$ is acceleration due to gravity, $a$ and $d$ are the radius and equilibrium submergence of the cylinder and $s$ is its heave amplitude, which is assumed to be small. Equations \eqref{BVP} are solved in conjunction with a prescribed incident wave, a Sommerfeld radiation condition as $r\to\infty$, a cube-root singularity condition at the submerged edge of the cylinder, and a frequency-domain equation of motion of the WEC, which is given as
\begin{equation}
    -\omega^2\rho d\pi a^2s=\mathrm{i}\omega\mu s-k_ss-\rho g \pi a^2 s+\mathrm{i}\omega\rho\int_{-\pi}^\pi \int_0^a\phi(r,\theta,-d)r\upd r\upd\theta.\label{eom_FD}
\end{equation}
The terms on the right hand side of the above equation correspond to the damping, spring, hydrostatic and hydrodynamic forces, respectively, with $\rho$, $\mu$ and $k_s$ being the fluid density, damping coefficient and spring coefficient, respectively. By rotational symmetry, the solution to \eqref{BVP} can be expressed as
\begin{equation}
    \phi(r,\theta,z) = \sum_{n=-\infty}^\infty \phi_n(r,z)\e^{\upi n\theta}
\end{equation}
where the functions $\phi_n$ have a piecewise definition depending on whether a point $(r,z)$ is beneath the WEC (i.e., $r<a$) or not (i.e., $r>a$), that is
\begin{equation}\label{matching}
    \phi_n(r,z)=\begin{cases}
        \varphi_n(r,z)&r>a,\quad z\in(-H,0)\\
        \chi_n(r,z)&r<a,\quad z\in(-H,-d).
    \end{cases}
\end{equation}
The functions $\varphi_n$ and $\chi_n$ are obtained using separation of variables as
\begin{subequations}
\begin{align}
\varphi_n(r,z)&=\sum_{m=0}^\infty\psi_m(z)\left(A_{mn}\frac{J_n(k_mr)}{J_n^\prime(k_ma)}+B_{mn}\frac{H_n^{(1)}(k_mr)}{H_n^{(1)\prime}(k_ma)}\right)\label{exterior_field}\\
\chi_n(r,z)&=\begin{cases} C_{0n}\frac{r^{|n|}}{|n|a^{|n|-1}}+\sum_{m=1}^\infty \tilde{\psi_m}(z)C_{mn}\frac{I_n(\kappa_mr)}{I_n^\prime(\kappa_m a)}&n\neq 0\\
C_{00}+\sum_{m=1}^\infty \tilde{\psi_m}(z)C_{m0}\frac{I_0(\kappa_mr)}{I_0^\prime(\kappa_m a)}-\frac{\mathrm{i}\omega s}{2h}\left((z+H)^2-\frac{r^2}{2}\right)&n=0\end{cases}\label{chi_def}
\end{align}
\end{subequations}
where $J_n$, $H_n^{(1)}$ and $I_n$ denote the Bessel, Hankel and modified Bessel functions of the first kind of order $n$, respectively. The quantities $k_m$ are the solutions to the dispersion relation $k\tanh(kH)=\omega^2/g$, with $k_0\in\mathbb{R}^+$ being the wavenumber associated with propagating waves. Additionally, $\kappa_m=m\pi/h$, where $h=H-d$. The vertical eigenfunctions are given by $\psi_m(z)=\beta_m^{-1/2}\cosh(k_m(z+H))$, where $\beta_m=\sinh(2k_mH)/4k_mH+\tfrac{1}{2}$, and $\tilde{\psi_m}(z)=\sqrt{2}\cos(\kappa_m(z+H))$. The term proportional to $s$ in \eqref{chi_def} is a particular solution for the inhomogeneous boundary condition \eqref{piston_dyn_BC} \citep{Yeung1981}.

Equation \eqref{matching} gives rise to a matching problem for each $n$ between the interior region $r<a$ and the exterior region $r>a$, as $\phi_n$ must be continuously differentiable at $r=a$, namely
\begin{subequations}\label{matching2}
\begin{align}
\varphi_n(a,z)&=\chi_n(a,z)\label{continuous_potential}\\
\partial_r\varphi_n(a,z)&=\partial_r\chi_n(a,z),\label{continuous_derivative}
\end{align}
\end{subequations}
for $z\in(-H,-d)$. It follows that the matching problem for a given $n$ is to determine the coefficients $B_{mn}$ and $C_{mn}$ in terms of the coefficients $A_{mn}$, which are characterised by a known incident wave. The matching problem for $\phi_0$ must be solved in tandem with the equation of motion \eqref{eom_FD}, and the WEC heave amplitude $s$ is determined from this problem alone. Conversely, for $n\neq 0$, the modes $\phi_n$ are uncoupled from the motion of the WEC, and are equivalent to the modes of diffraction around a fixed cylinder.

\subsection{Solution using an integral equation/Galerkin method}\label{integral_eq_subsec}
While early solutions of problems of wave scattering by truncated cylinders used the eigenfunction matching method \citep{Garrett_1971,Yeung1981}, here we follow \citet{LI2019952} and use a singularity respecting integral equation/Galerkin method. Such methods typically provide more accuracy for the same computing time \citep{jmse13030398}. Although we require the solution to the matching problem \eqref{matching2} for all $n$ is solved, here we only show the calculation for the case when $n=0$, as this is the mode which is coupled to the WEC dynamics. The cases for $n\neq 0$ are similar, though importantly they are independent of the vertical motion of the WEC. More complete explanations of the method are given by \citet{porterThesis1995} and \citet{kanoria1999water}.

The condition that $\partial_r\phi_0$ is continuous at $r=a$ \eqref{continuous_derivative}, together with \eqref{barrel}, gives
\begin{equation}
u(z)=\partial_r\varphi_0(a,z)=\begin{cases}
    0&z>-d\\
    \partial_r\chi_0(a,z)&z<-d,
\end{cases}
\end{equation}
where we have introduced an auxiliary function $u$. In terms of the eigenfunction expansions, we have
\begin{subequations}
\begin{align}
u(z)&=\sum_{m=0}^\infty\psi_m(z)(A_{m0}+B_{m0})\\
&=\frac{\upi\omega sa}{2h}+\sum_{m=1}^\infty\tilde{\psi_m}(z)C_{m0}.\label{u_expression_coeffs_2}
\end{align} 
\end{subequations}
Using the orthogonality of the vertical eigenfunctions, namely
\begin{equation}
\int_{-H}^{0}\psi_m(z)\psi_n(z)\upd z=H\delta_{mn}\quad\text{and}\quad\int_{-H}^{-d}\tilde{\psi_m}(z)\tilde{\psi_n}(z)\upd z=h\delta_{mn},
\end{equation}
the following expressions for the unknown coefficients are obtained in terms of $u$:
\begin{subequations}\label{coefficient_integral_expressions}
\begin{align}
B_{m0}&=\frac{1}{H}\int_{-H}^{-d}\psi_m(\xi)u(\xi)\upd\xi-A_{m0}&m\geq 0\\
C_{m0}&=\frac{1}{h}\int_{-H}^{-d}\tilde{\psi_m}(\xi)u(\xi)\upd\xi&m>0.\label{C_m0_equation}
\end{align}
\end{subequations}
Note that in \eqref{C_m0_equation}, there is no dependence on $s$ since
\begin{equation}
\int_{-H}^{-d}\tilde{\psi_m}(z)\upd z=0,
\end{equation}
for all $m>0$. The requirement that $\phi_0$ itself is also continuous at $r=a$ \eqref{continuous_potential}, combined with the expressions for the coefficients \eqref{coefficient_integral_expressions}, eventually gives rise to the following integral equation:
\begin{equation}\label{integral_eq}
\int_{-H}^{-d}\mathcal{K}(z,\xi)u(\xi)\upd\xi+C_{00}-\frac{\upi\omega s}{2h}\left((z+H)^2-\frac{a^2}{2}\right)=\mathcal{F}(z),
\end{equation}
where
\begin{align*}
\mathcal{K}&=\frac{1}{h}\sum_{m=1}^\infty \frac{I_0(\kappa_ma)}{I_0^\prime(\kappa_ma)}\tilde{\psi_m}(z)\tilde{\psi_m}(\xi)-\frac{1}{H}\sum_{m=0}^\infty \frac{H_0^{(1)}(k_ma)}{H_0^{(1)\prime}(k_ma)}\psi_m(z)\psi_m(\xi)\\
\mathcal{F}(z)&=\sum_{m=0}^\infty A_{m0}\left(\frac{J_0(k_m a)}{J_0^\prime(k_ma)}- \frac{H_0^{(1)}(k_ma)}{H_0^{(1)\prime}(k_ma)}\right)\psi_m(z).
\end{align*}

To obtain a numerical solution of the integral equation \eqref{integral_eq}, the auxiliary function is expanded in a specified basis as
\begin{equation}
u(z)=\sum_{j=0}^{N_{\mathrm{aux}}}c_jv_j(z),
\end{equation}
where $N_{\mathrm{aux}}$ is a truncation parameter. Substituting this expression into \eqref{integral_eq} and imposing Galerkin orthogonality (i.e., the residual error is orthogonal to the basis functions $v_p$ for $0\leq p \leq N_{\mathrm{aux}}$) gives a system of $N_{\mathrm{aux}}+1$ equations and $N_{\mathrm{aux}}+3$ unknowns---these being the auxiliary coefficients $c_j$, the coefficient $C_{00}$ and the piston amplitude $s$. In order to determine the system, two additional equations are required. The first equation is obtained by integrating \eqref{u_expression_coeffs_2} over $(-H,-d)$, giving
\begin{equation}\label{second_relation}
\int_{-H}^{-d}u(\xi)\upd\xi=\frac{\upi\omega s a}{2},
\end{equation}
where the left hand side can be expressed in terms of the auxiliary coefficients $c_j$. The second equation is provided by the equation of motion \eqref{eom_FD}, where we note that the integral on the right hand side is independent of $\phi_n$ for $n\neq 0$, since the terms $e^{\upi n \theta}$ integrate to zero.

The resulting system of equations has a matrix representation of the form
\begin{equation}\label{matrix_n_0}
\begin{bmatrix}
    K&\boldsymbol{\gamma}^1&\boldsymbol{\gamma}^2\\(\boldsymbol{\gamma}^1)^\intercal&0&-\upi\omega a/2\\(\boldsymbol{\gamma}^3)^\intercal &\upi\omega\rho\pi a^2&\square
\end{bmatrix}
\begin{bmatrix}
    \mathbf{c}\\C_{00}\\s
\end{bmatrix}=
\begin{bmatrix}
    \mathbf{F}\\0\\0
\end{bmatrix}
\end{equation}
where the $N_{\mathrm{aux}}+1$-dimensional matrix $K$ and vectors $\boldsymbol{\gamma}^1$, $\boldsymbol{\gamma}^2$ and $\mathbf{F}$ are derived from the integral equation \eqref{integral_eq} and Galerkin orthogonality, and have entries
\begin{subequations}
\begin{align}
    K_{pj} &= \int_{-H}^{-d}\int_{-H}^{-d}v_z(\xi)\mathcal{K}(z,\xi)v_j(\xi)\upd\xi\upd z\\
    \gamma_1^p&=\int_{-H}^{-d}v_p(z)\upd z\\
    \gamma^2_p&=\frac{-\upi\omega}{2h}\int_{-H}^{-d}\left((z+H)^2-\frac{a^2}{2}\right)v_p(z)\upd z\\
    F_p &=\int_{-H}^{-d}\mathcal{F}(z)v_p(z)\upd z,
\end{align}
\end{subequations}
where we note that $F_p$ depends linearly on the incident wave coefficients $A_{m0}$. The entries of the vector $\mathbf{c}$ are the auxiliary coefficients $c_j$. The second row of the matrix expression \eqref{matrix_n_0} is derived from \eqref{second_relation}, and the third row is derived from the equation of motion \eqref{eom_FD}. For brevity, the entries of the $N_{\mathrm{aux}}+1$-dimensional vector $\boldsymbol{\gamma}^3$ are not stated---we merely note that they express the dependence of the integral in \eqref{eom_FD} on the auxiliary coefficients $c_j$ via the coefficients $C_{m0}$, for $m\geq 1$. The number $\square$ is the only element of the matrix in \eqref{matrix_n_0} which depends on the WEC spring-damper properties $k_s$ and $\mu$. If the radius $a$ and depth $d$ are held fixed, the remaining entries of the matrix in \eqref{matrix_n_0} can be precomputed, which rapidly accelerates computation. Once the system in \eqref{matrix_n_0} is solved, the scattered wave coefficients $B_{m0}$ are readily obtained from the auxiliary coefficients $c_j$.

In order to capture the cube root singularity of $\nabla\phi_n$ at the submerged corner $(r,z)=(a,-d)$, the auxiliary basis is taken to consist of weighted Gegenbauer polynomials of the form \citep{kanoria1999water}
\begin{equation}
    v_j(z)=\frac{2^{7/6}\Gamma(1/6)(2j)!}{\pi\Gamma(2j+1/3)h^{1/3}}(h^2-(z+H)^2)^{-1/3}C_{2m}^{1/6}\left(\frac{z+H}{h}\right)
\end{equation}

\subsection{The T-matrix for a WEC}\label{T_matrix_subsec}
Solutions of the matching problem for a range of $n$ are used to obtain a $T$-matrix expression of the form
\begin{equation}\label{T-matrix}
    \begin{bmatrix}
    t_{-N_W}&&&&\\
    &\ddots&&&\\
    &&t_0&&\\
    &&&\ddots&\\
    &&&&t_{N_W}
    \end{bmatrix}\begin{bmatrix}A_{0,-N_W}\\ \vdots \\ A_{0,0}\\\vdots\\A_{0,N_W}\end{bmatrix}=\begin{bmatrix}B_{0,-N_W}\\ \vdots \\ B_{0,0}\\\vdots\\B_{0,N_W}\end{bmatrix}
\end{equation}
which relates incident wave amplitudes to scattered wave amplitudes. The truncation parameter $N_W$ is obtained from Wiscombe's formula \citep{Wiscombe:80}, which empirically describes the number of angular modes required for convergence as a function of the nondimensional WEC radius $k_0a$. The $T$-matrix given in \eqref{T-matrix} is based on the wide spacing approximation, i.e., evanescent modes (terms for $m>0$ in \eqref{exterior_field}) are not included. The underlying assumption of the wide spacing approximation is that a negligible amount of energy is transmitted by evanescent modes between WECs in the subsequent multiple scattering problem. Note that the $T$-matrix is diagonal due to rotational symmetry. Moreover, the only entry of the $T$-matrix which depends on the spring and damping coefficients is $t_0$; provided $a$ and $d$ are fixed, the remaining entries can be precomputed which rapidly accelerates computation.

For subsequent energy absorption calculations, we also require a way of calculating the heave amplitude of the WEC $s$ from the incident coefficients $A_{mn}$. By linearity, rotational symmetry, and an assumption that incident evanescent modes are negligible in the excitation of the WEC (i.e., the wide spacing approximation) we note that there is a scalar $s_{\mathrm{rel}}$ satisfying $s_{\mathrm{rel}}A_{0,0} = s$, although the details of its derivation are not presented here for brevity.

\subsection{Scattering by multiple WECs}

Next, we consider the multiple scattering problem of $N$ WECs centred at $(x_j,y_j)$ for $1\leq j\leq N$, each having constant radius $a$ and equilibrium submergence $d$ but with varying spring and damping coefficients $k_{s,j}$ and $\mu_j$, respectively. The exterior field can be written as a superposition of the incident wave and the waves scattered by all WECs as \citep{martin2006}
\begin{equation}
\phi \approx \phi_{\mathrm{inc}}+\sum_{j=1}^N\phi_{\mathrm{sc}}^{(j)}.
\end{equation}
We approximate the wave scattered by cylinder $j$ using the wide spacing approximation and Wiscombe's truncation formula as
\begin{equation}
\phi_{\mathrm{sc}}^{(j)}(r_j,\theta_j,z)=\psi_0(z)\sum_{n=-N_W}^{N_W}B_{0n}^{(j)}\frac{H_n^{(1)}(k_0 r_j)}{H_n^{(1)\prime}(k_0a)}e^{\upi n\theta},
\end{equation}
where $(r_j,\theta_j,z)$ are local cylindrical coordinate systems  centred at the WEC central axes. They are related to the global cartesian coordinate system by $x=x_j+r_j\cos\theta_j$, and $y=y_j+r_j\sin\theta_j$. The system is solved using the self consistent theory of multiple scattering \citep{Kagemoto_Yue_1986,PETER_MEYLAN_2004,martin2006,montielMeylanHawkins}. The theory results in a system of equations of the form
\begin{equation}\label{Graf_system}
    (I-T_{\mathrm{Array}}S_{\mathrm{Array}})\begin{bmatrix}
        \mathbf{B}^{(1)}\\\vdots\\\mathbf{B}^{(N)}
    \end{bmatrix}=\mathbf{F}
\end{equation}
where $I$ is the $N(2N_W+1)$-dimensional identity matrix and the block-diagonal matrix $T_{\mathrm{Array}}$ is defined as
\begin{equation}
    T_{\mathrm{Array}}=\begin{bmatrix}
        T^{(1)}&&\\
        &\ddots&\\
        &&T^{(N)}
    \end{bmatrix}
\end{equation}
where $T^{(j)}$ is the $T$-matrix of the form \eqref{T-matrix} associated with the $j$th scatterer. Moreover, $S_{\mathrm{Array}}$ is a matrix arising from Graf's addition theorem depending on the frequency and the relative positions of the WECs, but not on their properties. When solving the wave energy park layout optimisation problem described in \textsection\ref{optimisation_sec}, evaluation of the entries of $S_{\mathrm{Array}}$ is one of the most expensive parts of the computation---it cannot be precomputed as the layout varies during the solution procedure. The entries of the vectors $\mathbf{B}^{(j)}$ are $B_{0n}^{(j)}$, i.e., the coefficients of the wave scattered by the $j$th WEC. Lastly, the forcing vector $\mathbf{F}$ depends on the incident wave $\phi_{\mathrm{inc}}$. Full details of the derivation of \eqref{Graf_system} are provided by \citet{martin2006}. We note that once the scattered wave amplitudes $B_{0n}^{(j)}$ are known, it is straightforward to compute $s_j$, namely, the complex heave amplitude of each WEC, using quantities analogous to $s_{\mathrm{rel}}$ (introduced in \textsection\ref{T_matrix_subsec}) for each WEC.

\subsection{Energy absorption}
We consider the array to be subject to irregular plane waves propagating in the positive $x$ direction. Assuming a simple model of a fully developed sea generated by wind-wave interactions, we describe the sea state using the Pierson-Moskowitz spectrum \citep{pierson1964}
\begin{equation}
    S(\omega)=\frac{c_1 g^2}{\omega^5}\exp\left(-\frac{c_2 g^2}{\omega^4 H_s^2}\right)
\end{equation}
in which $c_1=8.1\times 10^{-3}$ and $c_2=3.24\times10^{-2}$ are fixed constants and the significant wave height $H_s$ is a free parameter, which we take to be $H_s=2$\,m throughout this paper. The average rate of energy absorption by the array in this sea is given by
\begin{equation} \label{objective_fun}
    P_S(\mathbf{X}) = \int_0^\infty P(\mathbf{X},\omega)\sqrt{2S(\omega)\upd\omega},
\end{equation}
where $\mathbf{X}$ is a vector containing the parameters of the array, i.e. the coordinates of the WECs and the parameters of their spring-damper systems. The functions $P(\mathbf{X},\omega)$ represent the average rate of energy absorption by the array when subjected to a unit-amplitude monochromatic plane wave of frequency $\omega$, travelling in the positive $x$ direction. They can be calculated directly from the piston amplitudes as
\begin{equation}\label{near_field_energy}
    P(\mathbf{X},\omega)=\sum_{j=1}^N \tfrac{1}{2}\omega^2\mu_j|s_j|^2
\end{equation}
where $\mu_j$ and $s_j$ are the damping coefficient and computed amplitude of the $j$th piston, respectively. The functions $P(\mathbf{X},\omega)$ can also be calculated at the far field via a generalised optical theorem \citep{Mei2005}
\begin{equation}\label{far_field_energy}
    P(\mathbf{X},\omega)=\omega\rho H\left[\frac{1}{\pi}\int_{-\pi}^\pi |\mathcal{B}(\theta)|^2d\theta-\frac{2\sqrt{\beta_0}g}{\omega\cosh(k_0H)}\mathrm{Im}(A^*\mathcal{B}(0))\right]
\end{equation}
where the far field function is defined as
\begin{equation}
\mathcal{B}(\theta)=\sum_{j=1}^N\exp(-\mathrm{i}k_0R_{0j}\cos(\varphi_{0j}-\theta))\sum_{n=-\infty}^\infty \frac{(-\mathrm{i})^nB_{0n}^{(j)}}{H_n^{(1)\prime}(k_0a)}e^{\mathrm{i}n\theta},
\end{equation}
where $(R_{0j},\varphi_{0j})$ are the polar coordinates of $(x_j,y_j)$ with respect the global polar coordinate system $(r,\theta)$. Agreement between \eqref{near_field_energy} and \eqref{far_field_energy}, which is equivalent to conservation of energy, is used to validate our computations.

During our numerical computations, the integral in \eqref{objective_fun} \citep[of which the technical details are discussed in][]{ochi1998} is approximated using the following quadrature rule
\begin{equation}
P_S(\mathbf{X})\approx \sum_{j=1}^{100} P(\mathbf{X},\omega_j)\sqrt{2S(\omega_j)\Delta\omega}
\end{equation}
where $\omega_j=\omega_1+(j-1)\Delta\omega$ are evenly spaced quadrature points, which are chosen to span an interval over which the spectral density function $S(\omega)$ is non-negligible. In the case where the significant wave height is $H_s=2$\,m, the quadrature points are chosen so that $\omega_1=0.4$\,s$^{-1}$ and $\omega_{100}=4$\,s$^{-1}$ (i.e., wave periods between $1.57$\,s and $15.71$\,s).

\section{Optimisation}\label{optimisation_sec}

We set up an optimisation problem that consists in finding $\mathbf{X}_{\mathrm{opt}}$ which maximises the objective function $f(\mathbf{X})$ subject to constraints on the parameters in $\mathbf{X}$. The parameters subject to optimisation are the WEC coordinates $(x_j,y_j)$, and their spring and damping coefficients $k_{s,j}$ and $\mu_j$, respectively. The objective function $f(\mathbf{X})$ is proportional to $P_S(\mathbf{X})$ except when the distance between any two WECs becomes small---in such situations, a penalty is applied to prevent the assumption of wide spacing from being violated. In particular
\begin{equation}
    f(\mathbf{X})=\begin{cases}
        f_+(\mathbf{X})&R_{\mathrm{min}}>4a\\
        \theta f_-(\mathbf{X})+(1-\theta)f_+(\mathbf{X})&3a<R_{\mathrm{min}}<4a\\
        f_-(\mathbf{X})&R_{\mathrm{min}}<3a,
    \end{cases}
\end{equation}
where $R_{\mathrm{min}}=\min_{i\neq j}R_{ij}$ is the minimum distance between any two pairs of WECs. Moreover, $f_+=P_S(\mathbf{X})/P_{S,\mathrm{MAX}}$ and $f_-=R_{\mathrm{min}}/(4a)-1$ are the positive reward and negative penalty functions, respectively, and $\theta=4-R_{\mathrm{min}}/a$ is a convex sum parameter. The scalar $P_{S,\mathrm{MAX}}$ is an estimate of the upper bound of the energy absorbed by the array, given by
\begin{equation}
P_{S,\mathrm{MAX}}=N\int_0^\infty \frac{\rho g\omega}{4k_0^2}\left(1+\frac{2k_0H}{\sinh(2k_0H)}\right)\sqrt{2S(\omega)\upd\omega},
\end{equation}
in which we note that $k_0$ depends on $\omega$ through the dispersion relation. The above estimated bound assumes each WEC is maximally efficient at all frequencies \citep{Mei2005} and is uncoupled from all other WECs. While we make no attempt to prove that this is an upper bound of $P_S(\mathbf{X})$, we note that it was not exceeded during our computations. The resulting objective function $f(\mathbf{X})$ is expected to be a continuous function of $\mathbf{X}$ taking values in $[-1,1]$.

For all $1\leq j\leq N$, the WEC coordinates $(x_j,y_j)$ are constrained to the bounding box $0\leq x_j\leq D_x$ and $0\leq y_j\leq D_y$. To simplify computation, we additionally require that (i) $x_1=0$ and (ii) $x_1\leq x_2\leq...\leq x_N$. Note that these do not restrict the optimal solutions due to (i) a translation argument and (ii) a permutation argument. Note that condition (i) reduces the number of parameters by $1$, meaning that $\mathbf{X}$ is  $(4N-1)$-dimensional. The spring coefficients are restricted to $|k_{s,j}|\leq 10^5$\,kg\,s$^{-2}$. Note that this constraint permits negative spring coefficients, which is not a new idea in wave energy conversion \citep{zhang2016oscillating,TODALSHAUG201668,en11092362,Westcott_Bennetts_Sergiienko_Cazzolato_2024}. Springs with negative stiffness decrease the resonant frequency of an individual WEC. This could alternatively be achieved by increasing $d$, i.e., the equilibrium submergence depth of a WEC, although varying $k_s$ is more computationally efficient than varying $d$ because it allows more quantities to be precomputed (as discussed in \textsection\ref{integral_eq_subsec}). The damping coefficients are restricted to $0\leq \mu_j\leq 2\times 10^3$\,kg\,s$^{-1}$. The bounds on $k_{s,j}$ and $\mu_j$ are sufficiently non restrictive so that an individual WEC can achieve its theoretical optimal capture width $1/k_0$ across the frequency interval considered in this study $\omega\in[0.4,4]$\,s$^{-1}$. Note that the aforementioned capture width is the length of a wave crest carrying the same amount of energy as the WEC absorbs---the reader is referred to \citet{Mei2005} for further details on this quantity and its theoretical bounds for axisymmetric devices.

The results presented in this paper were obtained using MATLAB's built-in genetic algorithm \verb|ga| with the default settings. A sequential optimisation algorithm motivated by those used in previous studies of rainbow absorption was also considered \citep{jimenez2017,Wilks2022,Westcott_Bennetts_Sergiienko_Cazzolato_2024}, but this yielded poorer optimal solutions than those obtained using the genetic algorithm . Specifically, the sequential algorithm consisted of an iteration, in which each step begins with the optimal array of $N-1$ WECs, adds a new WEC in a random location to obtain an initial guess $\mathbf{X}_0$, then performs a local search of the parameter space beginning at $\mathbf{X}_0$ using the MATLAB algorithm \verb|fmincon|. As they are better, we only present results obtained using the genetic algorithm in this paper.

\section{Results}\label{results_sec}

\begin{figure}
    \centering
    \includegraphics[width=\textwidth]{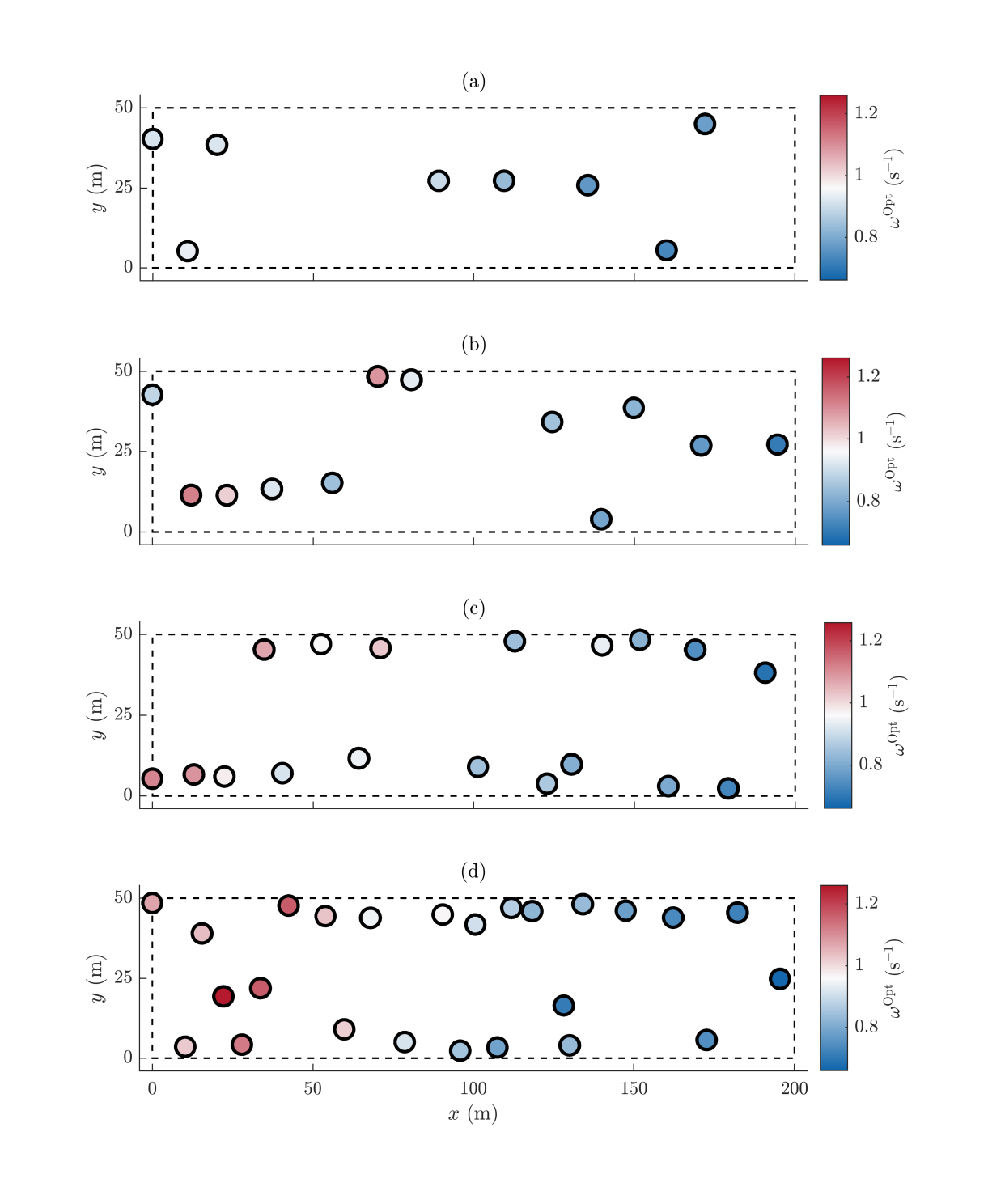}
    \caption{Optimised configurations of an array of (a) $N=8$, (b) $N=12$, (c) $N=18$ and (d) $N=25$ WECs found using the genetic algorithm, when constrained to a rectangular bounding box $D_x=200$\,m and $D_y=50$\,m (dashed line). The WECs are drawn as circles, whose radii are not drawn to scale. The colour of the WECs indicates $\omega^{\mathrm{Opt}}$, i.e.\ the frequency at which that WEC would optimally absorb energy if uncoupled from all other WECs. The direction of wave propagation is from left to right.}
    \label{fig:optimal_rectangular_arrays}
\end{figure}

\begin{figure}
    \centering
    \includegraphics[width=\linewidth]{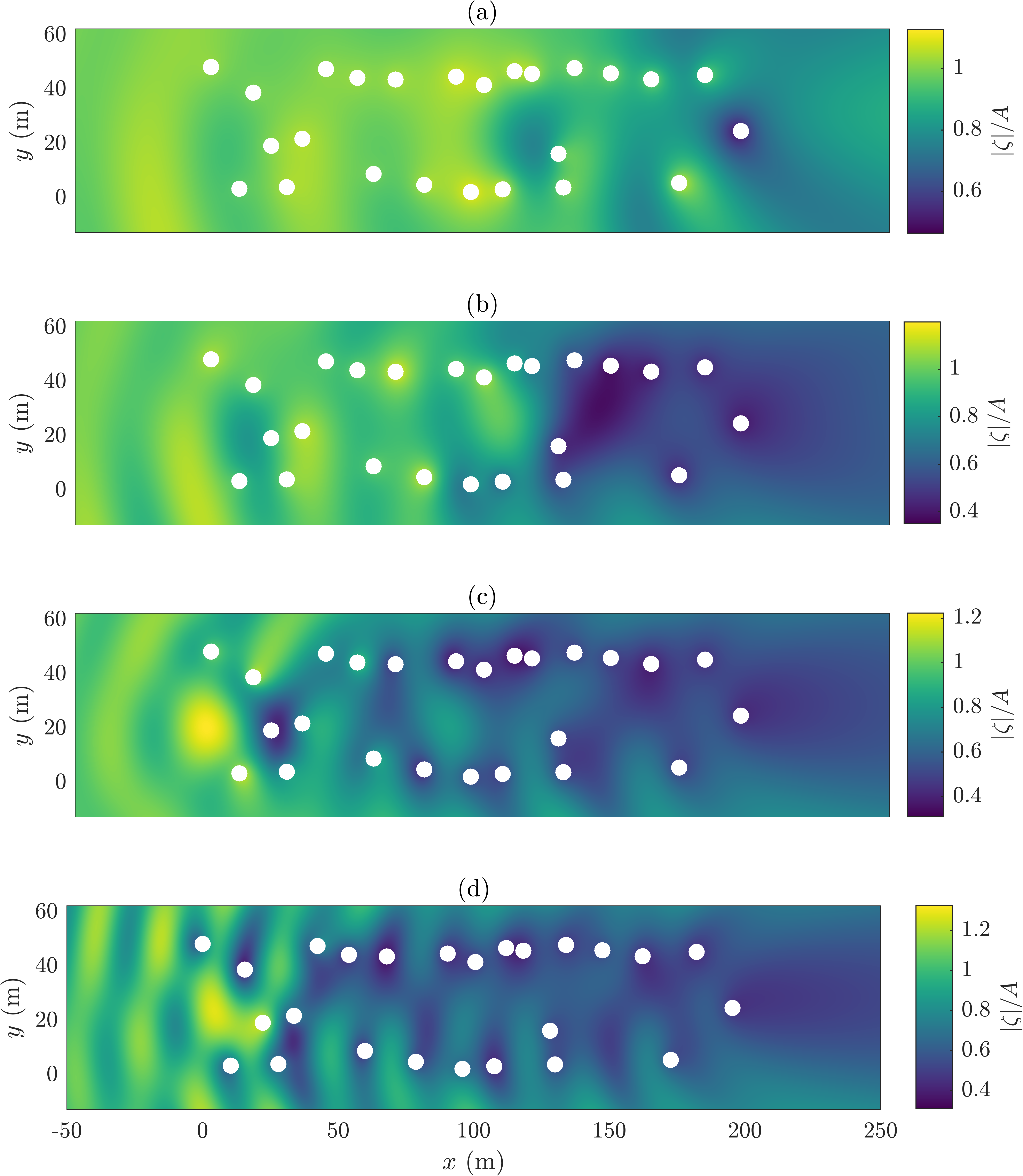}
    \caption{Non-dimensionalised free surface elevation resulting from wave scattering by the array of 25 WECs optimised for the rectangular bounding box $D_x=200$\,m and $D_y=50$\,m, at (a) $\omega=0.7$\,s$^{-1}$, (b) $\omega=0.85$\,s$^{-1}$, (c) $\omega=1$\,s$^{-1}$ and (d) $\omega=1.15$\,s$^{-1}$. The WECs are plotted as white circles (not to scale). The non-dimensionalised free surface is defined as $|\zeta(x,y)|/A$, where $\zeta(x,y)$ is the free surface resulting from excitation by a plane wave with amplitude $A$.}
    \label{fig:rectangle_free_surface}
\end{figure}

Figure \ref{fig:optimal_rectangular_arrays} shows optimised array configurations for a rectangular bounding box $D_x=200$\,m and $D_y=50$\,m, for an increasing number of WECs. We observe that WECs which are most efficient at higher frequencies occur towards the front of the array, whereas WECs which are most efficient at lower frequencies occur towards the rear of the array. While there does appear to be a grading of the WECs' peak absorption efficiency frequencies across the array, this grading is not monotonic. As the number of WECs $N$ increases, we observe the gradual onset of two parallel pseudo-line arrays approximately along $y=0$\,m and $y=D_x$. We emphasise that these are not true line arrays as the WECs are not collinear. We hypothesise that these pseudo--line arrays arise in the optimal configurations due to the rainbow absorption effect. In particular, we propose that waves analogous to Rayleigh--Bloch waves propagate along these pseudo-line arrays and gradually slow down as the frequency of the local resonance decreases. The local energy amplification resulting from this subsequently results in efficient absorption. Figure \ref{fig:rectangle_free_surface} shows the free surface elevation resulting from wave interaction with one of these optimal arrays at four different frequencies. Qualitatively, we observe that lower frequency waves amplify further towards the right of the array, which supports the conclusion that rainbow absorption underpins absorption by this array.

\begin{figure}
    \centering
    \includegraphics[width=\textwidth]{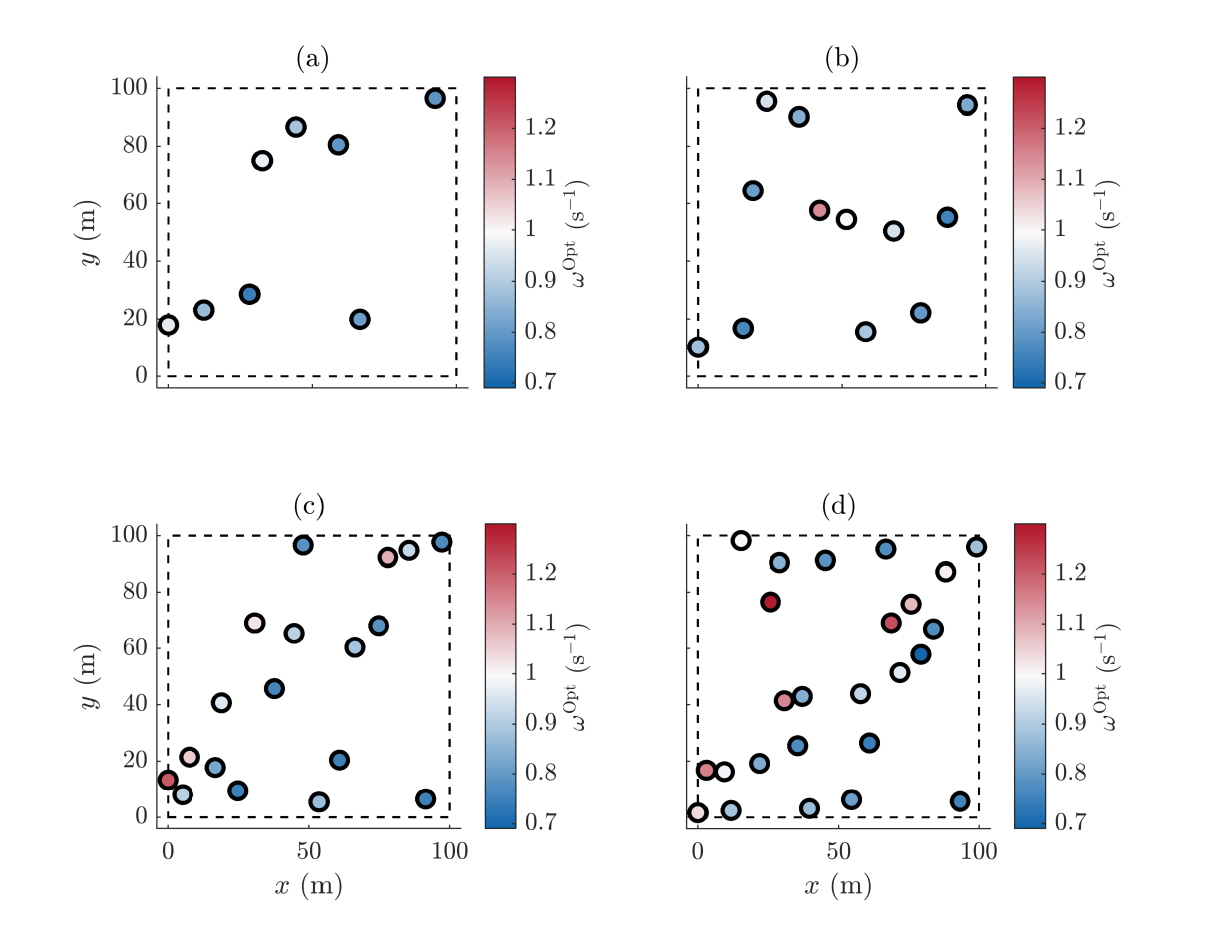}
\caption{Optimised configurations of an array of (a) $N=8$, (b) $N=12$, (c) $N=18$ and (d) $N=25$ WECs found using the genetic algorithm, when constrained to a rectangular bounding box $D_x=D_y=100$\,m (dashed line). The WECs are drawn as circles, whose radii are not drawn to scale. The colour of the WECs indicates $\omega^{\mathrm{Opt}}$, i.e.\ the frequency at which that WEC would optimally absorb energy if uncoupled from all other WECs. The direction of wave propagation is from left to right.}
    \label{fig:optimal_square_arrays}
\end{figure}

\begin{figure}
    \centering
    \includegraphics[width=\linewidth]{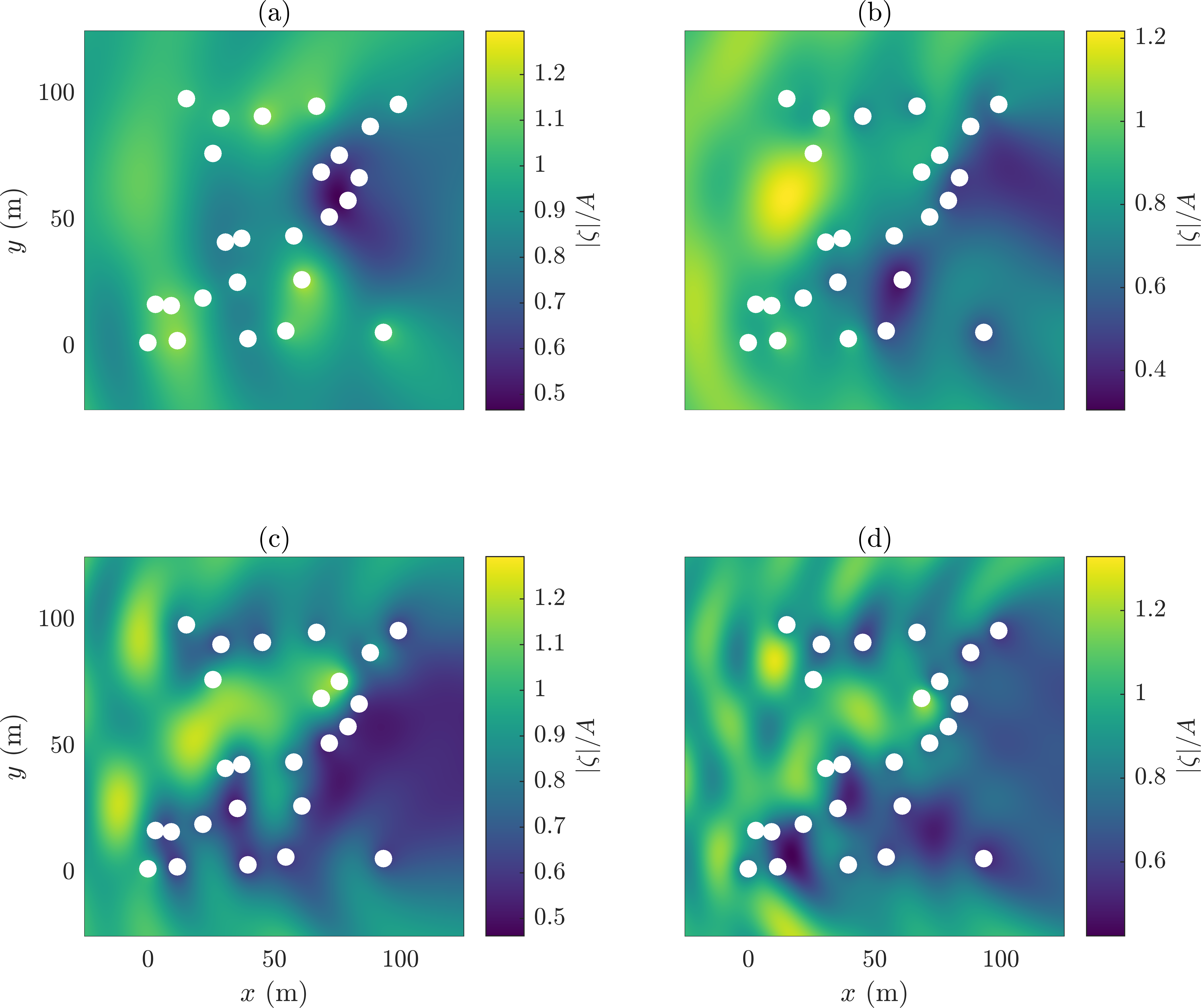}
    \caption{Non-dimensionalised free surface elevation resulting from wave scattering by the array of 25 WECs optimised for the square bounding box $D_x=D_y=100$\,m, at (a) $\omega=0.7$\,s$^{-1}$, (b) $\omega=0.85$\,s$^{-1}$, (c) $\omega=1$\,s$^{-1}$ and (d) $\omega=1.15$\,s$^{-1}$. The WECs are plotted as white circles (not to scale). The non-dimensionalised free surface is defined as $|\zeta(x,y)|/A$, where $\zeta(x,y)$ is the free surface resulting from excitation by a plane wave with amplitude $A$.}
    \label{fig:square_free_surface}
\end{figure}

\begin{figure}
    \centering
    \includegraphics[width=\linewidth]{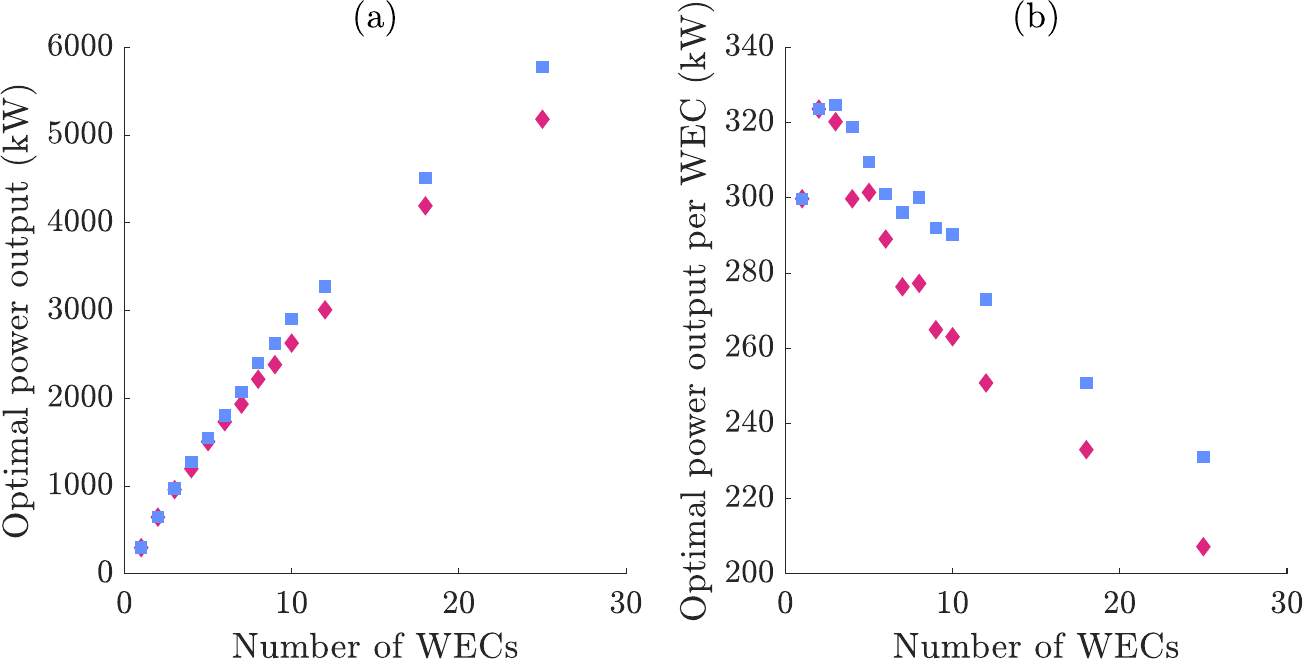}
    \caption{(a) Rate of energy absorption by optimised arrays of WECs $P(\mathbf{X}_{\mathrm{opt}})$ as a function of the array size $N$. Values for the rectangular bounding box problem ($D_x=200$\,m and $D_y=50$\,m) are marked with magenta diamonds and those for the square bounding box problem ($D_x=D_y=100$\,m) are marked with blue squares. (b) As for panel (a), except the vertical axis displays the rate of energy absorption per WEC in the optimised arrays, i.e., $P(\mathbf{X}_{\mathrm{opt}})/N$.}
    \label{fig:effect_of_N}
\end{figure}

Figure \ref{fig:optimal_square_arrays} shows the optimised array configurations for a square bounding box $D_x=D_y=100$\,m. While in some cases we observe local grading of the WECs' peak absorption efficiency frequencies, this is not the case globally as the array layouts are very irregular. Thus, interactions between WECs in square-bounded arrays are presumably much more complex than can be understood through the rainbow reflection effect alone. Figure \ref{fig:square_free_surface} shows the free surface elevation resulting from wave interaction with one of these optimal arrays at four different frequencies. In contrast to the rectangular array in Figure \ref{fig:rectangle_free_surface}, there is little qualitative evidence of rainbow absorption. While there appears to be some indication that low-frequency waves propagate further than high-frequency waves in the panel $y<30$\,m, this is not conclusive. We also remark that square-bounded arrays absorb more energy than rectangle-bounded arrays with the same number of WECs, as shown in Figure \ref{fig:effect_of_N}(a). Presumably, this is because the square bounding box is twice as narrow in the $x$-direction, meaning that WECs at the rear of the array are subject to weaker shadowing effects.

\begin{figure}
    \centering
    \includegraphics[width=0.6\linewidth]{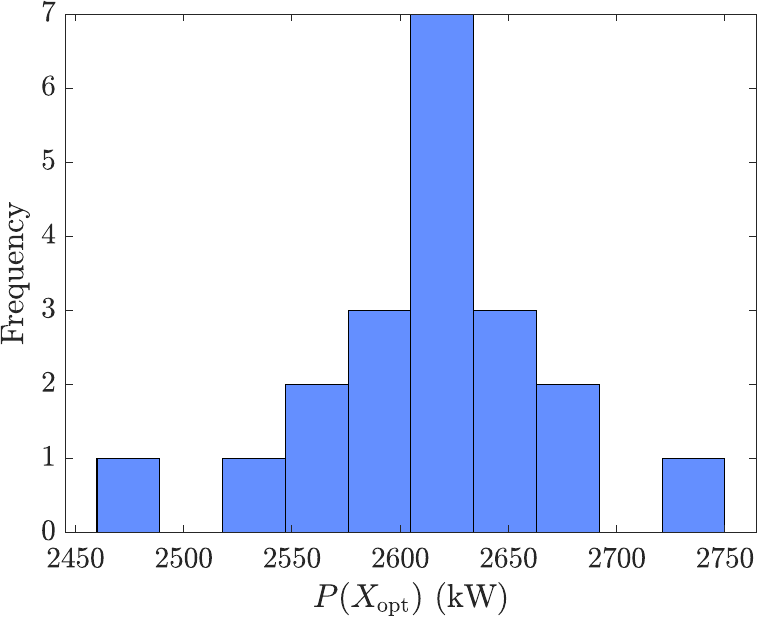}
    \caption{Histogram showing the rate of energy absorption of optimal arrays found from $20$ independent evaluations of the genetic algorithm. Each array consists of $10$ WECs bound to the rectangular region ($D_x=200$\,m and $D_y=50$\,m).}
    \label{fig:GA_error}
\end{figure}

Because the objective function is not convex, it is important to consider whether the optimal arrays in Figures \ref{fig:optimal_rectangular_arrays} and \ref{fig:optimal_square_arrays} are global optima or, if they are only local optima, how significantly they deviate from the global optima. In Figure \ref{fig:GA_error}, a histogram is shown for $20$ independent realisations of the genetic algorithm in the case $N=20$, $D_x=200$\,m and $D_y=50$\,m. There is a considerable spread in the range of power take-off values, with the best array outperforming the median by $5\%$. Thus, the optimal arrays shown in Figures \ref{fig:optimal_rectangular_arrays} and \ref{fig:optimal_square_arrays} should only be considered as local optima. We note that a repeated evaluation of the genetic algorithm was not conducted in general in this article due to the computational expense of these evaluations, particularly for large $N$.

Figure \ref{fig:effect_of_N} illustrates the effect of the number of WECs $N$ on the power take-off by the optimal arrays. We observe that for both square-bounded and rectangle-bounded arrays, the addition of each successive WEC increases the total power take-off by the array. However, the rate of power take-off per WEC is predominantly decreasing as a function of $N$, indicating that there are diminishing returns in adding more WECs to the array. 

\begin{figure}
    \centering
    \includegraphics[width=\linewidth]{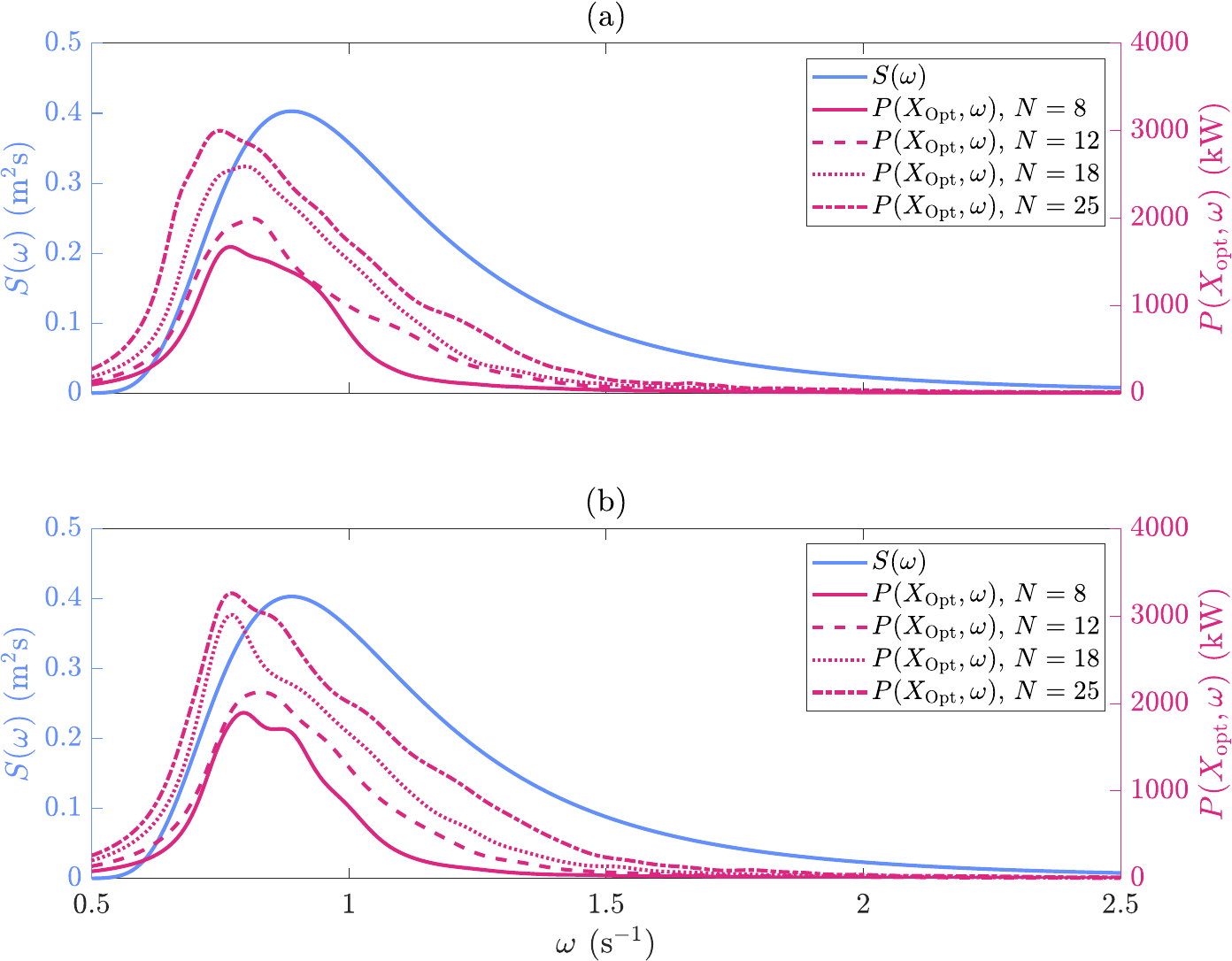}
    \caption{Power take-off spectra $P(\mathbf{X}_{\mathrm{opt}},\omega)$ for the optimal arrays constrained to the (a) rectangular and (b) square bounding boxes, as displayed in Figures \ref{fig:optimal_rectangular_arrays} and \ref{fig:optimal_square_arrays}, respectively. The spectral density function $S(\omega)$, which is a Pierson-Moskowitz spectrum with significant wave height $H_s=2$\,m, is also shown for comparison.}
    \label{fig:power_take-off_spectra}
\end{figure}

Figure \ref{fig:power_take-off_spectra} shows power take-off spectra $P(\mathbf{X}_{\mathrm{opt}},\omega)$ for the optimal arrays displayed in Figures \ref{fig:optimal_rectangular_arrays} and \ref{fig:optimal_square_arrays}. Increasing the number of WECs $N$ appears to confer increased power take-off at all frequencies, for both rectangle-bounded and square-bounded arrays. We also observe that the peaks of the power take-off spectra occur at lower frequencies than the peak of the spectral density function $S(\omega)$, presumably because lower frequency waves have a higher group velocity, and therefore transport energy into the array at a higher rate.

\section{Discussion}\label{discussion_sec}
This paper has developed a model of a wave energy park with an arbitrary WEC layout. In the model, each WEC consists of a heaving truncated cylinder that is coupled to a spring-damper system which models power take-off. The layout and device parameters were optimised subject to constraints using a genetic algorithm. The optimisation procedure yielded wave energy park configurations which are local maximisers of energy absorption in a unidirectional, irregular sea. When constrained to a rectangular bounding box that is elongated in the direction of wave propagation, the optimal layout consist of graded pseudo-line arrays when the number of WECs is sufficiently large. Moreover, low-frequency waves propagate further into the array than high-frequency waves, which is indicative of rainbow absorption. Our results were less conclusive in the case of arrays constrained to a square bounding box, where we suspect that phenomena other than rainbow absorption govern the interactions between WECs in the array.

A limitation of this study is the extent to which we have explored the parameter space of our model. One class of parameters that has not been considered is those governing the sea state, namely the significant wave height $H_s$ and the direction of the incident waves. Different choices of these parameters would change the objective function, thereby giving rise to different optimal arrays. While the effect of the sea state on the optimal arrays could be explored using a discrete collection of parameters, a more comprehensive understanding of this effect could be obtained by extending the optimisation problem to a multi-condition problem of the form
\begin{equation}
    \min_{\mathbf{X}\in\Omega}f(\mathbf{X},\mathbf{C}) \quad\text{for all }\mathbf{C}\in\Phi.
\end{equation}
In the above, $\Omega$ is the decision space (here describing the array layout and device parameters) and $\Phi$ is the condition space describing parameters which are external to the optimisation problem (here describing the sea state). The task of a multi-condition optimisation problem is to find the optimal solution for all conditions, that is, to find an optimal vector $\mathbf{X}_{\mathbf{C}}\in\Omega$ for all values of $\mathbf{C}\in\Phi$. One avenue for future work is to optimise arrays of WECs using deep reinforcement learning-based algorithms which have recently been proposed for multi-condition optimisation tasks \citep{kim2022,balasooriya2024multi}.

A second limitation of this work is whether the optimal WEC spring and damping coefficients found in this study are realistic. Future studies could consider what parameter ranges can be engineered in prototype WECs and investigate the array configurations that arise under these restrictions. Another potential weakness is the use of a linearised kinematic condition at the submerged face of each WEC, which can permit unrealistically large heave amplitudes of floating bodies at resonance, particularly when their horizontal extent is much smaller than one wavelength.

Lastly, we mention the important question of whether the irregularity of the optimal arrays, particularly those bounded to the square region, is a feature of optimal arrays or a limitation of the optimisation procedure. In support of the former hypothesis, we tangentially mention the work of \citet{Chaplain_2020}, who studied graded arrays of energy-harvesting rods in the context of elastic waves, in which the rods were grouped into triangles in each unit cell. The authors illustrated that the introduction of asymmetry in the unit cell greatly increased the potential for energy harvesting, as it decoupled the wave carrying incident energy into the array from its counterpart carrying reflected energy out of the array. In the same vein, a graded Su--Schrieffer--Heeger metawedge of energy harvesting rods, in which the spacing of the rods is alternating instead of uniform, yielded considerably greater energy amplifications than a conventionally graded structure due to topologically protected edge states \citep{PhysRevApplied.14.054035}. In light of these studies, it is perhaps unsurprising that the optimal arrays found in this paper are irregular, although the precise nature of the interaction in these arrays is unknown.

\subsection*{Acknowledgements}
BW, DSB, TJ, CW and SC are supported by the Australian Research Council’s Linkage Projects funding scheme (LP190100378) and Australian Government Research Training Program Scholarships to DSB and TJ. BW and MHM are supported by the Australian Research Council's Discovery Projects funding scheme (DP240102104).  FM is supported by the Royal Society Te Ap\={a}rangi (Marsden Fund projects 20-UOO-173) and New Zealand's Antarctic Science Platform (project ANTA1801).

\bibliographystyle{plainnat}
\bibliography{bibfile}

\begin{thebibliography}{39}
\providecommand{\natexlab}[1]{#1}
\providecommand{\url}[1]{\texttt{#1}}
\expandafter\ifx\csname urlstyle\endcsname\relax
  \providecommand{\doi}[1]{doi: #1}\else
  \providecommand{\doi}{doi: \begingroup \urlstyle{rm}\Url}\fi

\bibitem[Arbon{\`e}s et~al.(2018)Arbon{\`e}s, Sergiienko, Ding, Krause, Igel, and Wagner]{sergiienko2018}
D{\'i}dac~Rodr{\'i}guez Arbon{\`e}s, Nataliia~Y. Sergiienko, Boyin Ding, Oswin Krause, Christian Igel, and Markus Wagner.
\newblock Sparse incomplete lu-decomposition for wave farm designs under realistic conditions.
\newblock In Anne Auger, Carlos~M. Fonseca, Nuno Louren{\c{c}}o, Penousal Machado, Lu{\'i}s Paquete, and Darrell Whitley, editors, \emph{Parallel Problem Solving from Nature -- PPSN XV}, pages 512--524, Cham, 2018. Springer International Publishing.
\newblock ISBN 978-3-319-99253-2.

\bibitem[Balasooriya et~al.(2024)Balasooriya, Blair, Wheeler, and Chalup]{balasooriya2024multi}
Dasun~Shalila Balasooriya, Alan Blair, Craig Wheeler, and Stephan Chalup.
\newblock Multi-condition multi-objective airfoil shape optimisation using deep reinforcement learning compared to genetic algorithms.
\newblock In \emph{International Conference on Optimization, Learning Algorithms and Applications}, pages 243--258. Springer Nature Switzerland, 2024.
\newblock \doi{https://doi.org/10.1007/978-3-031-77432-4_17}.

\bibitem[Bergstr{\"o}m and G{\"o}teman(2024)]{bergstrom2024comprehensive}
K~Bergstr{\"o}m and M~G{\"o}teman.
\newblock Comprehensive multi-objective optimisation of wave power parks.
\newblock In \emph{Innovations in Renewable Energies Offshore}, pages 295--305. CRC Press, 2024.
\newblock \doi{10.1201/9781003558859-33}.

\bibitem[Chaplain et~al.(2020{\natexlab{a}})Chaplain, Pajer, De~Ponti, and Craster]{Chaplain_2020}
G~J Chaplain, Daniel Pajer, Jacopo~M De~Ponti, and R~V Craster.
\newblock Delineating rainbow reflection and trapping with applications for energy harvesting.
\newblock \emph{New Journal of Physics}, 22\penalty0 (6):\penalty0 063024, jun 2020{\natexlab{a}}.
\newblock \doi{10.1088/1367-2630/ab8cae}.
\newblock URL \url{https://dx.doi.org/10.1088/1367-2630/ab8cae}.

\bibitem[Chaplain et~al.(2020{\natexlab{b}})Chaplain, De~Ponti, Aguzzi, Colombi, and Craster]{PhysRevApplied.14.054035}
Gregory~J. Chaplain, Jacopo~M. De~Ponti, Giulia Aguzzi, Andrea Colombi, and Richard~V. Craster.
\newblock Topological rainbow trapping for elastic energy harvesting in graded {Su-Schrieffer-Heeger Systems}.
\newblock \emph{Phys. Rev. Appl.}, 14:\penalty0 054035, Nov 2020{\natexlab{b}}.
\newblock \doi{10.1103/PhysRevApplied.14.054035}.
\newblock URL \url{https://link.aps.org/doi/10.1103/PhysRevApplied.14.054035}.

\bibitem[Child(2011)]{child2011configuration}
Benjamin Frederick~Martin Child.
\newblock \emph{On the configuration of arrays of floating wave energy converters}.
\newblock PhD thesis, University of Edinburgh, 2011.

\bibitem[Child and Venugopal(2010)]{CHILD20101402}
B.F.M. Child and V.~Venugopal.
\newblock Optimal configurations of wave energy device arrays.
\newblock \emph{Ocean Engineering}, 37\penalty0 (16):\penalty0 1402--1417, 2010.
\newblock ISSN 0029-8018.
\newblock \doi{https://doi.org/10.1016/j.oceaneng.2010.06.010}.
\newblock URL \url{https://www.sciencedirect.com/science/article/pii/S0029801810001447}.

\bibitem[de~O.~Falcão(2010)]{FALCAO2010899}
António~F. de~O.~Falcão.
\newblock Wave energy utilization: A review of the technologies.
\newblock \emph{Renewable and Sustainable Energy Reviews}, 14\penalty0 (3):\penalty0 899--918, 2010.
\newblock ISSN 1364-0321.
\newblock \doi{https://doi.org/10.1016/j.rser.2009.11.003}.
\newblock URL \url{https://www.sciencedirect.com/science/article/pii/S1364032109002652}.

\bibitem[Falnes(2007)]{FALNES2007185}
Johannes Falnes.
\newblock A review of wave-energy extraction.
\newblock \emph{Marine Structures}, 20\penalty0 (4):\penalty0 185--201, 2007.
\newblock ISSN 0951-8339.
\newblock \doi{https://doi.org/10.1016/j.marstruc.2007.09.001}.
\newblock URL \url{https://www.sciencedirect.com/science/article/pii/S0951833907000482}.

\bibitem[Garrett(1971)]{Garrett_1971}
C.~J.~R. Garrett.
\newblock Wave forces on a circular dock.
\newblock \emph{Journal of Fluid Mechanics}, 46\penalty0 (1):\penalty0 129–139, 1971.
\newblock \doi{10.1017/S0022112071000430}.

\bibitem[Giassi and Göteman(2018)]{GIASSI2018252}
Marianna Giassi and Malin Göteman.
\newblock Layout design of wave energy parks by a genetic algorithm.
\newblock \emph{Ocean Engineering}, 154:\penalty0 252--261, 2018.
\newblock ISSN 0029-8018.
\newblock \doi{https://doi.org/10.1016/j.oceaneng.2018.01.096}.
\newblock URL \url{https://www.sciencedirect.com/science/article/pii/S0029801818301045}.

\bibitem[Golbaz et~al.(2022)Golbaz, Asadi, Amini, Mehdipour, Nasiri, Etaati, Naeeni, Neshat, Mirjalili, and Gandomi]{GOLBAZ202215446}
Danial Golbaz, Rojin Asadi, Erfan Amini, Hossein Mehdipour, Mahdieh Nasiri, Bahareh Etaati, Seyed Taghi~Omid Naeeni, Mehdi Neshat, Seyedali Mirjalili, and Amir~H. Gandomi.
\newblock Layout and design optimization of ocean wave energy converters: A scoping review of state-of-the-art canonical, hybrid, cooperative, and combinatorial optimization methods.
\newblock \emph{Energy Reports}, 8:\penalty0 15446--15479, 2022.
\newblock ISSN 2352-4847.
\newblock \doi{https://doi.org/10.1016/j.egyr.2022.10.403}.
\newblock URL \url{https://www.sciencedirect.com/science/article/pii/S235248472202337X}.

\bibitem[G{\"o}teman et~al.(2020)G{\"o}teman, Giassi, Engstr{\"o}m, and Isberg]{goteman2020advances}
Malin G{\"o}teman, Marianna Giassi, Jens Engstr{\"o}m, and Jan Isberg.
\newblock Advances and challenges in wave energy park optimization—a review.
\newblock \emph{Frontiers in Energy Research}, 8:\penalty0 26, 2020.
\newblock \doi{10.3389/fenrg.2020.00026}.

\bibitem[Guo and Ringwood(2021)]{guo2021}
Bingyong Guo and John~V. Ringwood.
\newblock A review of wave energy technology from a research and commercial perspective.
\newblock \emph{IET Renewable Power Generation}, 15\penalty0 (14):\penalty0 3065--3090, 2021.
\newblock \doi{https://doi.org/10.1049/rpg2.12302}.

\bibitem[Jim{\'e}nez et~al.(2017)Jim{\'e}nez, Romero-Garc{\'\i}a, Pagneux, and Groby]{jimenez2017}
No{\'e} Jim{\'e}nez, Vicent Romero-Garc{\'\i}a, Vincent Pagneux, and Jean-Philippe Groby.
\newblock Rainbow-trapping absorbers: Broadband, perfect and asymmetric sound absorption by subwavelength panels for transmission problems.
\newblock \emph{Scientific reports}, 7\penalty0 (1):\penalty0 1--12, 2017.

\bibitem[Kagemoto and Yue(1986)]{Kagemoto_Yue_1986}
Hiroshi Kagemoto and Dick K.~P. Yue.
\newblock Interactions among multiple three-dimensional bodies in water waves: an exact algebraic method.
\newblock \emph{Journal of Fluid Mechanics}, 166:\penalty0 189–209, 1986.
\newblock \doi{10.1017/S0022112086000101}.

\bibitem[Kanoria et~al.(1999)Kanoria, Dolai, and Mandal]{kanoria1999water}
Mridula Kanoria, DP~Dolai, and BN~Mandal.
\newblock Water-wave scattering by thick vertical barriers.
\newblock \emph{Journal of engineering mathematics}, 35:\penalty0 361--384, 1999.
\newblock \doi{10.1023/A:1004392622976}.

\bibitem[Kim et~al.(2022)Kim, Kim, and You]{kim2022}
Sejin Kim, Innyoung Kim, and Donghyun You.
\newblock Multi-condition multi-objective optimization using deep reinforcement learning.
\newblock \emph{Journal of Computational Physics}, 462:\penalty0 111263, 2022.
\newblock \doi{10.1016/j.jcp.2022.111263}.

\bibitem[Li and Liu(2019)]{LI2019952}
{Ai-jun} Li and Yong Liu.
\newblock New analytical solutions to water wave diffraction by vertical truncated cylinders.
\newblock \emph{International Journal of Naval Architecture and Ocean Engineering}, 11\penalty0 (2):\penalty0 952--969, 2019.
\newblock ISSN 2092-6782.
\newblock \doi{10.1016/j.ijnaoe.2019.04.006}.

\bibitem[Linton and McIver(2001)]{Linton2001}
C.~M. Linton and P.~McIver.
\newblock \emph{{Handbook of mathematical techniques for wave/structure interactions}}.
\newblock Chapman {\&} Hall, 2001.
\newblock \doi{10.1201/9781420036060}.

\bibitem[Martin(2006)]{martin2006}
Paul~A Martin.
\newblock \emph{Multiple scattering: interaction of time-harmonic waves with N obstacles}.
\newblock Number 107 in Encyclopedia of Mathematics and its Applications. Cambridge University Press, 2006.

\bibitem[McGuinness(2018)]{mcguinness2018hydrodynamic}
Justin~PL McGuinness.
\newblock \emph{Hydrodynamic optimisation of an array of wave-power devices}.
\newblock PhD thesis, University College Cork, 2018.

\bibitem[Mei et~al.(2005)Mei, Stiassnie, and Yue]{Mei2005}
Chiang~C Mei, M~Stiassnie, and Dick K.-P. Yue.
\newblock \emph{{Theory and applications of ocean surface waves Part 1: Linear aspects}}.
\newblock World Scientific, 2005.

\bibitem[Montiel et~al.(2024)Montiel, Meylan, and Hawkins]{montielMeylanHawkins}
F.~Montiel, M.~H. Meylan, and S.~C. Hawkins.
\newblock Scattering kernel of an array of floating ice floes: application to water wave transport in the marginal ice zone.
\newblock \emph{Proceedings of the Royal Society A: Mathematical, Physical and Engineering Sciences}, 480\penalty0 (2282):\penalty0 20230633, 2024.
\newblock \doi{10.1098/rspa.2023.0633}.

\bibitem[Neshat et~al.(2020)Neshat, Alexander, Sergiienko, and Wagner]{NESHAT2020100744}
Mehdi Neshat, Bradley Alexander, Nataliia~Y. Sergiienko, and Markus Wagner.
\newblock New insights into position optimisation of wave energy converters using hybrid local search.
\newblock \emph{Swarm and Evolutionary Computation}, 59:\penalty0 100744, 2020.
\newblock ISSN 2210-6502.
\newblock \doi{https://doi.org/10.1016/j.swevo.2020.100744}.
\newblock URL \url{https://www.sciencedirect.com/science/article/pii/S2210650220303977}.

\bibitem[Ochi(1998)]{ochi1998}
Michel~K. Ochi.
\newblock \emph{Ocean Waves: The Stochastic Approach}.
\newblock Number~6 in Cambridge Ocean Technology Series. Cambridge University Press, 1998.

\bibitem[Peter and Meylan(2004)]{PETER_MEYLAN_2004}
Malte~A. Peter and Michael~H. Meylan.
\newblock Infinite-depth interaction theory for arbitrary floating bodies applied to wave forcing of ice floes.
\newblock \emph{Journal of Fluid Mechanics}, 500:\penalty0 145–167, 2004.
\newblock \doi{10.1017/S0022112003007092}.

\bibitem[Pierson~Jr and Moskowitz(1964)]{pierson1964}
Willard~J Pierson~Jr and Lionel Moskowitz.
\newblock A proposed spectral form for fully developed wind seas based on the similarity theory of {SA Kitaigorodskii}.
\newblock \emph{Journal of geophysical research}, 69\penalty0 (24):\penalty0 5181--5190, 1964.

\bibitem[Porter(1995)]{porterThesis1995}
Richard Porter.
\newblock \emph{Complementary Methods and Bounds in Linear Water Waves}.
\newblock PhD thesis, University of Bristol, 1995.

\bibitem[Sheng(2019)]{SHENG2019482}
Wanan Sheng.
\newblock Wave energy conversion and hydrodynamics modelling technologies: A review.
\newblock \emph{Renewable and Sustainable Energy Reviews}, 109:\penalty0 482--498, 2019.
\newblock ISSN 1364-0321.
\newblock \doi{https://doi.org/10.1016/j.rser.2019.04.030}.
\newblock URL \url{https://www.sciencedirect.com/science/article/pii/S1364032119302424}.

\bibitem[Teixeira-Duarte et~al.(2022)Teixeira-Duarte, Clemente, Giannini, Rosa-Santos, and Taveira-Pinto]{TEIXEIRADUARTE2022112513}
Felipe Teixeira-Duarte, Daniel Clemente, Gianmaria Giannini, Paulo Rosa-Santos, and Francisco Taveira-Pinto.
\newblock Review on layout optimization strategies of offshore parks for wave energy converters.
\newblock \emph{Renewable and Sustainable Energy Reviews}, 163:\penalty0 112513, 2022.
\newblock ISSN 1364-0321.
\newblock \doi{https://doi.org/10.1016/j.rser.2022.112513}.
\newblock URL \url{https://www.sciencedirect.com/science/article/pii/S1364032122004178}.

\bibitem[Todalshaug et~al.(2016)Todalshaug, Ásgeirsson, Hjálmarsson, Maillet, Möller, Pires, Guérinel, and Lopes]{TODALSHAUG201668}
Jørgen~Hals Todalshaug, Gunnar~Steinn Ásgeirsson, Eysteinn Hjálmarsson, Jéromine Maillet, Patrik Möller, Pedro Pires, Matthieu Guérinel, and Miguel Lopes.
\newblock Tank testing of an inherently phase-controlled wave energy converter.
\newblock \emph{International Journal of Marine Energy}, 15:\penalty0 68--84, 2016.
\newblock ISSN 2214-1669.
\newblock \doi{https://doi.org/10.1016/j.ijome.2016.04.007}.
\newblock URL \url{https://www.sciencedirect.com/science/article/pii/S2214166916300182}.
\newblock Selected Papers from the European Wave and Tidal Energy Conference 2015, Nante, France.

\bibitem[Têtu et~al.(2018)Têtu, Ferri, Kramer, and Todalshaug]{en11092362}
Amélie Têtu, Francesco Ferri, Morten~Bech Kramer, and Jørgen~Hals Todalshaug.
\newblock Physical and mathematical modeling of a wave energy converter equipped with a negative spring mechanism for phase control.
\newblock \emph{Energies}, 11\penalty0 (9), 2018.
\newblock ISSN 1996-1073.
\newblock \doi{10.3390/en11092362}.
\newblock URL \url{https://www.mdpi.com/1996-1073/11/9/2362}.

\bibitem[Westcott et~al.(2024)Westcott, Bennetts, Sergiienko, and Cazzolato]{Westcott_Bennetts_Sergiienko_Cazzolato_2024}
Amy-Rose Westcott, Luke~G. Bennetts, Nataliia~Y. Sergiienko, and Benjamin~S. Cazzolato.
\newblock Broadband near-perfect capture of water wave energy by an array of heaving buoy wave energy converters.
\newblock \emph{Journal of Fluid Mechanics}, 998:\penalty0 A5, 2024.
\newblock \doi{10.1017/jfm.2024.819}.

\bibitem[Wilks and Meylan(2025)]{jmse13030398}
Ben Wilks and Michael~H. Meylan.
\newblock A numerical comparison of eigenfunction matching and singularity-respecting galerkin approximation methods for linear water wave scattering.
\newblock \emph{Journal of Marine Science and Engineering}, 13\penalty0 (3), 2025.
\newblock \doi{10.3390/jmse13030398}.

\bibitem[Wilks et~al.(2022)Wilks, Montiel, and Wakes]{Wilks2022}
Ben Wilks, Fabien Montiel, and Sarah Wakes.
\newblock Rainbow reflection and broadband energy absorption of water waves by graded arrays of vertical barriers.
\newblock \emph{Journal of Fluid Mechanics}, 941, 2022.

\bibitem[Wiscombe(1980)]{Wiscombe:80}
W.~J. Wiscombe.
\newblock Improved {M}ie scattering algorithms.
\newblock \emph{Appl. Opt.}, 19\penalty0 (9):\penalty0 1505--1509, May 1980.
\newblock \doi{10.1364/AO.19.001505}.

\bibitem[Yeung(1981)]{Yeung1981}
Ronald~W. Yeung.
\newblock {Added mass and damping of a vertical cylinder in finite-depth waters}.
\newblock \emph{Applied Ocean Research}, 3\penalty0 (3):\penalty0 119--133, July 1981.
\newblock ISSN 0141-1187.
\newblock \doi{10.1016/0141-1187(81)90101-2}.

\bibitem[Zhang et~al.(2016)Zhang, Yang, and Xiao]{zhang2016oscillating}
Xian-tao Zhang, Jian-min Yang, and Long-fei Xiao.
\newblock An oscillating wave energy converter with nonlinear snap-through power-take-off systems in regular waves.
\newblock \emph{China Ocean Engineering}, 30\penalty0 (4):\penalty0 565--580, 2016.
\newblock \doi{10.1007/s13344-016-0035-5}.

\end{thebibliography}

\end{document}